\documentclass[11pt]{article}
\renewcommand{\baselinestretch}{1.1}
\usepackage{fullpage}
\usepackage[authoryear]{natbib}
\usepackage{hyperref}
\usepackage{amssymb,amsfonts,amsmath}
\usepackage{helvet}

\usepackage{amsmath}
\usepackage{amssymb}

\usepackage{graphicx}

\usepackage{cite}

\usepackage{enumerate}
\newcommand{\blank}[1]{}
\newcommand{\lt}{\left}
\newcommand{\rt}{\right}

\newcommand{\nn}{\nonumber}

\newcommand{\barF}{\overline{F}}
\newcommand{\barM}{\overline{M}}
\newcommand{\barn}{\overline{n}}

\blank{\usepackage{lineno}
\linenumbers
\newcommand*\patchAmsMathEnvironmentForLineno[1]{%
  \expandafter\let\csname old#1\expandafter\endcsname\csname #1\endcsname
  \expandafter\let\csname oldend#1\expandafter\endcsname\csname end#1\endcsname
  \renewenvironment{#1}%
     {\linenomath\csname old#1\endcsname}%
     {\csname oldend#1\endcsname\endlinenomath}}% 
\newcommand*\patchBothAmsMathEnvironmentsForLineno[1]{%
  \patchAmsMathEnvironmentForLineno{#1}%
  \patchAmsMathEnvironmentForLineno{#1*}}%
\AtBeginDocument{%
\patchBothAmsMathEnvironmentsForLineno{equation}%
\patchBothAmsMathEnvironmentsForLineno{align}%
\patchBothAmsMathEnvironmentsForLineno{flalign}%
\patchBothAmsMathEnvironmentsForLineno{alignat}%
\patchBothAmsMathEnvironmentsForLineno{gather}%
\patchBothAmsMathEnvironmentsForLineno{multline}%
}
}

\begin{document}
\begin{flushleft}
{\Large
\textbf{Reinterpreting Maximum Entropy in Ecology: a null hypothesis constrained by ecological mechanism}
}
% Insert Author names, affiliations and corresponding author email.
\\
James P. O'Dwyer $^{1}$, 
Andrew Rominger $^{2}$, 
Xiao Xiao $^3$
\\
\bf{1} Department of Plant Biology, University of Illinois, Urbana IL USA \\
\bf{2} Department of Environmental Science, Policy and Management, University of California, Berkeley, USA \\
\bf{3} School of Biology and Ecology, and Senator George J. Mitchell Center for Sustainability Solutions, University of Maine, Orono ME USA\\
\end{flushleft}

\textbf{Correspondence to be sent to:} \\Dr James P. O'Dwyer\\ Department of Plant Biology\\University of Illinois, Urbana IL 61801 \\ jodwyer@illinois.edu
\renewcommand{\baselinestretch}{1.1}
\normalsize

%%%%%%%%%%%%%%%%%%%%%%%%%%%%%%

%% For titles, only capitalize the first letter
%% \title{Almost sharp fronts for the surface quasi-geostrophic equation}

\section*{Abstract}
Simplified mechanistic models in ecology have been criticized for the fact that a good fit to data does not imply the mechanism is true: pattern does not equal process.  In parallel, the maximum entropy principle (MaxEnt) has been applied in ecology to make predictions constrained by just a handful of state variables, like total abundance or species richness.  But an outstanding question remains: what principle tells us which state variables to constrain? Here we attempt to solve both problems simultaneously, by translating a given set of mechanisms into the state variables to be used in MaxEnt, and then using this MaxEnt theory as a null model against which to compare mechanistic predictions. In particular, we identify the sufficient statistics needed to parametrize a given mechanistic model from data and use them as MaxEnt constraints. Our approach  isolates exactly what mechanism is telling us over and above the state variables alone. 
\pagebreak

\section*{Introduction}

Macroecology is the study of patterns of biodiversity aggregated across many species and individuals. These patterns encompass the distributions of organisms across space and time~\citep{preston1948commonness,preston1960time}, as well as multiple ways to quantify and measure biodiversity~\citep{morlon2009taking}. Macroecological patterns take surprisingly consistent, simple forms across many different taxonomic groups and distinct habitats~\citep{Rosenzweig1995}---for example, the distribution of rare and abundant species can be fitted using one of a handful of common distributions~\citep{preston1962canonical,May1975,mcGill2007species}.  This apparent universality, alongside the sense that it is driven by a combination of high diversity and large numbers~\citep{odwyer2014redqueen}, has led many ecologists to draw from statistical physics to understand and predict patterns of biodiversity~\citep{Harte1999,Odwyer2010}. Yet despite promising hints~\citep{mcgill2010towards,marquet2014theory}, we still currently lack the quantitative, overarching theoretical principles to explain how and why the forms of macroecological patterns are constrained.

\smallskip

The Maximum Entropy Theory of Ecology~\citep{harte2008maximum,harte2009biodiversity,harte2011maximum,harte2014maximum}, known as METE, has sought to fill this gap. The goal of this approach is to identify a probability distribution that can then be used to make predictions and tested against existing data. The principle of maximum entropy tells us that we can find a unique probability distribution that maximizes entropy, while constraining expectation values using the data we choose to feed into the algorithm. METE is very specific in terms of its data requirements: the theory prescribes a set of constraints based on intuitively important quantities such as the total number of individuals in a system, the total richness (of species or higher taxa \citep{harte2015}), and the total energy flux \citep{harte2011maximum}. While the specific values of these constraints will differ across ecological communities that are more or less diverse, productive, or populous, the theory posits that the appropriate constraints are identical for all systems. METE has been successful in a number of studies \citep{white2012characterizing, harte2014maximum}, and the cases where this prescription does work suggest that a large amount of the variance in macroecological patterns may indeed stem from statistical constraints \citep{harte2011maximum}.

\smallskip

However, the question arises, given that any data could be used to constrain the maximum entropy algorithm, why focus exclusively on the constraints proposed by METE?  This is particularly relevant given that METE does not universally succeed in predicting macroecological patterns \citep{newman2014, xiao2015strong}, potentially due to system-specific biological constraints being ignored, and the issue was raised in earlier METE papers~\citep{harte2008maximum}. Indeed we could constrain the maximum entropy algorithm with whatever data we know about a given ecological community, whether that is as specific as the number and spatial location of individuals of your favorite species, or as obscure as the skewness of the distribution of rare and abundant species. Whatever we think we know, the maximum entropy principle will then fill in the gaps, adding the least possible additional information.

\smallskip

For any mechanistic model with free, undetermined parameters, we always need to use some subset of our data to fit those parameters, before we can evaluate the performance of the model.  Our proposal is that we should use precisely the data necessary to fit mechanistic parameters as a constraint for a maximum entropy algorithm, and then use the corresponding MaxEnt predictions as a null model against which to compare mechanistic predictions.  If the mechanistic model then outperforms the corresponding MaxEnt distribution, then our choice of mechanism as modelers was successful.  If the model is outperformed by MaxEnt, we may as well not have modeled the mechanism at all---just using the subset of the data necessary to fit parameters and then maximizing entropy is a better approach.

\section{Mechanistic Models and Exponential Families}

Ecological theories often incorporate various forms of stochasticity, and hence make predictions for probability distributions rather than deterministic quantities.  These distributions can range over many questions and systems, from distributions of species abundance, to distributions of trait values.  For many (though by no means all) such theoretical predictions, these probability distributions turn out to belong to an exponential family, which means that the distribution is of the form:
\begin{align}
P(n) =A(\alpha)h(n)e^{-\alpha F(n)}\label{eq:generalexpfam}
\end{align}
for some functions $F(n)$, $h(n)$, $A(\alpha)$ and parameter $\alpha$.  In this language, the function $F(n)$ defines the ``family",  the base measure $h(n)$ and parameter $\alpha$ distinguish between different members of a given family, and the function $A(\alpha)$ ensures an appropriate normalization (i.e. probabilities sum or integrate to $=1$). We note that the support of the distribution depends on the ecological question and context, and we use the notation $n$ simply because several of our examples will involve discrete, positive species abundances. But this variable might equally represent the value of a continuous trait defined over a specified range, or most generally multiple variables of various types.   In general, exponential family distributions can take a diverse range of functional forms. These depend on the sufficient statistics (defined below), the base measure, and the number and type of variables, and include such common cases as the normal distribution, the gamma distribution, and the Pareto distribution, but also many other more general functions.

In the mechanistic models we will consider in this paper, the functions $F(n)$, $h(n)$, and $A(\alpha)$ will be essentially fixed by the theory, while $\alpha$ will be a parameter of the model.  In some situations, it might be possible to estimate this parameter from independently-gathered data. But in many cases, $\alpha$ will be a `free' parameter, something that ecologists must estimate using the data available. To be more specific, suppose this data takes the form of a series of $S$ independent observations of abundance, $\{n_i\}$.  Exponential families then have the property that measuring
\begin{align}
\barF=\frac{1}{S}\sum_{i=1}^{S} F(n_i)
\end{align}
is sufficient to compute a maximum likelihood estimate of the parameter $\alpha$.  We can see this explicitly by writing down the log likelihood of the parameter $\alpha$, given the $S$ independent observations, taking its derivative, and finding where this log likelihood is maximized:
\begin{align}
\frac{d}{d\alpha} \log \mathcal{L} & = S \frac{d \log A(\alpha)}{d\alpha} - \sum_{i=1}^S F(n_i)\nn\\
\Rightarrow & \lt.\frac{d \log A(\alpha)}{d\alpha}\rt|_{\alpha=\alpha_{\textrm{ML}}}  \ = \frac{1}{S}\sum_{i=1}^S F(n_i) = \barF\label{eq:generalmlest}
\end{align}
This yields an equation relating the maximum likelihood estimate of $\alpha$ to $\barF$.  In other words, in the case of a prediction belonging to an exponential family we need only a very precisely specified part of the data we've collected in order to fit the free parameter $\alpha$. $F(n)$ is therefore known as the sufficient statistic for this family of distributions, regardless of the form of the base measure, $h(n)$. On the other hand, while the data necessary to estimate $\alpha$ does not change with $h(n)$, different $h(n)$ do affect the value of the estimate of $\alpha$ through the normalization, $A(\alpha)$.

\subsection{Example: Species Abundances in a Neutral Model}

We now give an example: ecological neutral theory with no dispersal limitation~\citep{Hubbell2001,Volkov2003,Volkov2007,etienne2007neutral,Odwyer2012d, Rosindell2011}. In a neutral model, multiple species compete for a single resource, and interact completely symmetrically---no one species has a selective advantage over any other. From this assumption, it is possible to derive a neutral prediction for the distribution of species abundances.  To demonstrate this, we focus on a particular formulation of neutral theory known as the `non-zero-sum' model~\citep{haegeman2008relaxing}, and consider the probability $P_{\textrm{NT}}(n) $ that a species chosen at random has abundance $n$ (in a neutral world).  This distribution is the solution of a linear master equation, with effective birth and death rates, $b$ and $d$, that are the same for all species. The result is the well-known log series distribution:
\begin{equation}
P_{\textrm{NT}}(n) = \frac{1}{n\log(1/\nu)}{(1-\nu)^n}\label{eq:neutrallogseries}
\end{equation}
where $\nu = 1-\frac{b}{d}$, the difference between birth and death rates in units of the birth rate, and is constrained (by an assumption of constant community size) to be equal to the per capita speciation rate. The log series itself had been introduced much earlier by Fisher~\citep{fisher1943relation}, and has been fitted to many data sets as a phenomenological candidate for empirical species abundance distributions~\citep{white2012characterizing}, and so the appearance of the same distribution purely arising from drift was a promising early result for the neutral hypothesis (with the caveat that species abundance distributions had been successfully reproduced by numerous alternative models).

How do we determine the best choice of parameter $\nu$ for a given data set?  If we had an appropriate independent data set with information about the speciation process, this could allow an independent estimate  of $\nu$ for a given system.  We could then compare the form of Eq.~\eqref{eq:neutrallogseries} with the corresponding observed species abundance distribution. In practice, ecologists testing neutral theory have interpreted $\nu$ as a free parameter to be fitted using the species abundance data.  We use the notation $n_i$ to denote the abundance of species $i$, and $S$, the total number of species in a community, thus the total abundance is $\sum_{i=1}^{S}n_i$.  We can straightforwardly recast Eq.~\eqref{eq:neutrallogseries} in the canonical form of an exponential family, by defining the parameter $\alpha = -\log(1-\nu)$.  Using this parametrization,
\begin{equation}
P_{\textrm{NT}}(n) =-\frac{1}{\log\lt(1-e^{-\alpha}\rt)} \frac{1}{n}e^{-\alpha n}\label{eq:logseries}
\end{equation}
This is now in the same form as Eq.~\eqref{eq:generalexpfam}, with $F(n)=n$, $h(n)=1/n$, and the normalization $A(\alpha) = \lt(-\log\lt(1-e^{-\alpha}\rt)\rt)^{-1}$, which ensures that $\sum_{n=1}^{\infty} P(n) = 1$.  For this distribution, the sufficient statistic is clearly $n$, so that the data we need to make a maximum likelihood estimate of the free parameter $\alpha$ is just the mean abundance per species, $\sum_i n_i/S$. Applying the solution for maximum likelihood estimates in Eq.~\eqref{eq:generalmlest} to the case $F(n)=n$, and translating back  explicitly to the speciation rate, $\nu$, the maximum likelihood estimate for $\nu$ satisfies the following equation:
\begin{equation}
\frac{1-\nu}{\nu\log(1/\nu)}= \sum_{i=1}^{S} n_i/S = \barn. \label{eq:logseriesml}
\end{equation}
We can then use Eq.~\eqref{eq:neutrallogseries}, with parametrization determined by Eq.~\eqref{eq:logseriesml}, to compute any measure of goodness of fit, or likelihood, or comparison with alternative models, or whatever we wish---all using this point estimate of $\nu$, which in turn requires $\barn$.   

Before going on, let's summarize what our definition of mechanistic model does and does not do, and how the example above can be generalized.  First, we are assuming that the mechanistic model specifies a set of degrees of freedom (for example, species abundances, $n$, above), and that the model also leads to a solution for a probability distribution over these degrees of freedom, and moreover that this distribution belongs to an exponential family.  We are also considering distributions where there remain one or more 'free' parameters, that encode some or other aspect of the ecological mechanism, but are not fixed to a particular value by the model itself.  In the neutral example, the only free parameter is speciation rate, $\nu$.  In principle, we might be able to estimate this parameter independently of the species abundance data, or at least have some prior distribution on parameter values based on our knowledge of speciation processes.  In this paper though we are considering models and contexts where all parameters that \textit{can} be fixed independently have been, and the remaining free parameters must be estimated using a given data set.  It is this perspective that leads us to the sufficient statistics for this particular model.

This neutral model assumes that there are no selective forces, and that species abundances change due to ecological drift alone.  We might therefore think of the dominant driver as being demographic stochasticity. What happens if we change the neutral assumption? Alternative mechanistic hypotheses for the species abundance distribution can also result in predicted distributions belonging to an exponential family, but with different sufficient statistics than the neutral model.  For example, if species abundances are driven by a large number of successive multiplicative factors (for example due to environmental stochasiticity), the central limit theorem leads to a log normal distribution of species abundances, which belongs to an exponential family with quite different sufficient statistics: $\log(n)$ and $\log(n)^2$~\citep{May1975}.    On the other hand, there is a key caveat here---it is by no means true that all ecological probability distributions will belong to an exponential family.  A classic example of a distribution which does not is the Cauchy distribution, which has appeared in a variety of ecological contexts, including predictions of animal movement~\citep{benhamou2007many}. In summary, our approach can be easily applied to a range of ecological predictions, so long as the relevant probability distribution falls into an exponential family.

\section{The Maximum Entropy Principle}

In many cases, we can also make a prediction for a probability distribution  using the principle of maximum entropy~\citep{Jaynes1957information}, which we abbreviate as MaxEnt.  MaxEnt here is defined as an inference principle, where the idea is to find the probability distribution that has the maximum possible entropy consistent with a given set of constraints from observed data.  In this context, entropy is defined as:
\begin{equation}
H = - \sum_{n=1}^\infty P(n) \log P(n),
\end{equation}  
and is such that larger values of $H$ correspond (on average) to smaller amounts of information about the distribution $P(n)$ in any single observation.  In addition to finding the distribution $P(n)$ that maximizes entropy, MaxEnt also allows for input from a given data set, in terms of constraints of the form:
\begin{equation}
\sum_{n=1}^\infty P(n)F(n) = \sum_i F(n_i)/S = \barF
\end{equation}
In other words, the MaxEnt distribution can be constrained so that its prediction for the theoretical average value of $F(n)$ (evaluated using the distribution $P(n)$) is equal to the observed average, $\barF$, for a given quantity, $F(n)$, calculated using values $F(n_i)$ drawn from empirical data.   For a single constraint, the MaxEnt distribution for $P(n)$ is~\citep{Jaynes1957information}
\begin{align}
P_{\textrm{ME}}(n) = B\lt(\lambda\rt)e^{-\lambda F(n)}
\end{align}
where the parameter $\lambda$ is known as a Lagrange multiplier, and $B\lt(\lambda\rt)$ is a normalization that depends on the values of this Lagrange multipliers for a particular data set. The value of $\lambda$ is then determined using the form of the MaxEnt distribution and the measured values of $F(n_i)$, which reduces to the expression:
\begin{align}
\frac{d\log B\lt(\lambda\rt)}{d \lambda} = \barF\label{eq:fixlagrange}.
\end{align}
where $\barF = \sum_{i=1}^S F(n_i)/S $  is an estimator for the expectation value of the observable quantity $F(n)$.

There are then straightforward generalizations to multiple constraints and continuous variables~\citep{Jaynes1957information}. For example, given $J$ variables $x_j$ and a set of $K$ constraints, $F_{k}(\{x_j\})$, where $j$ runs from $1$ to $J$ and $k$ runs from $1$ to $K$, the MaxEnt prediction for $P(\{x_j\})$ is:
\begin{align}
P_{\textrm{ME}}(\{x_j\}) = B\lt(\{\lambda_k\}\rt)e^{-\sum_{k=1}^{K}\lambda_k F_k(\{x_j\})}
\end{align}
where values of each of the Lagrange multipliers $\lambda_j$ is detemined by
\begin{align}
\frac{\partial \log B\lt(\{\lambda_k\}\rt)}{\partial \lambda_j} = \bar{F_j}.
\end{align}

\begin{figure}
\setlength{\unitlength}{1cm}
\includegraphics[scale=0.64,clip=TRUE,trim= 0 80 0 50]{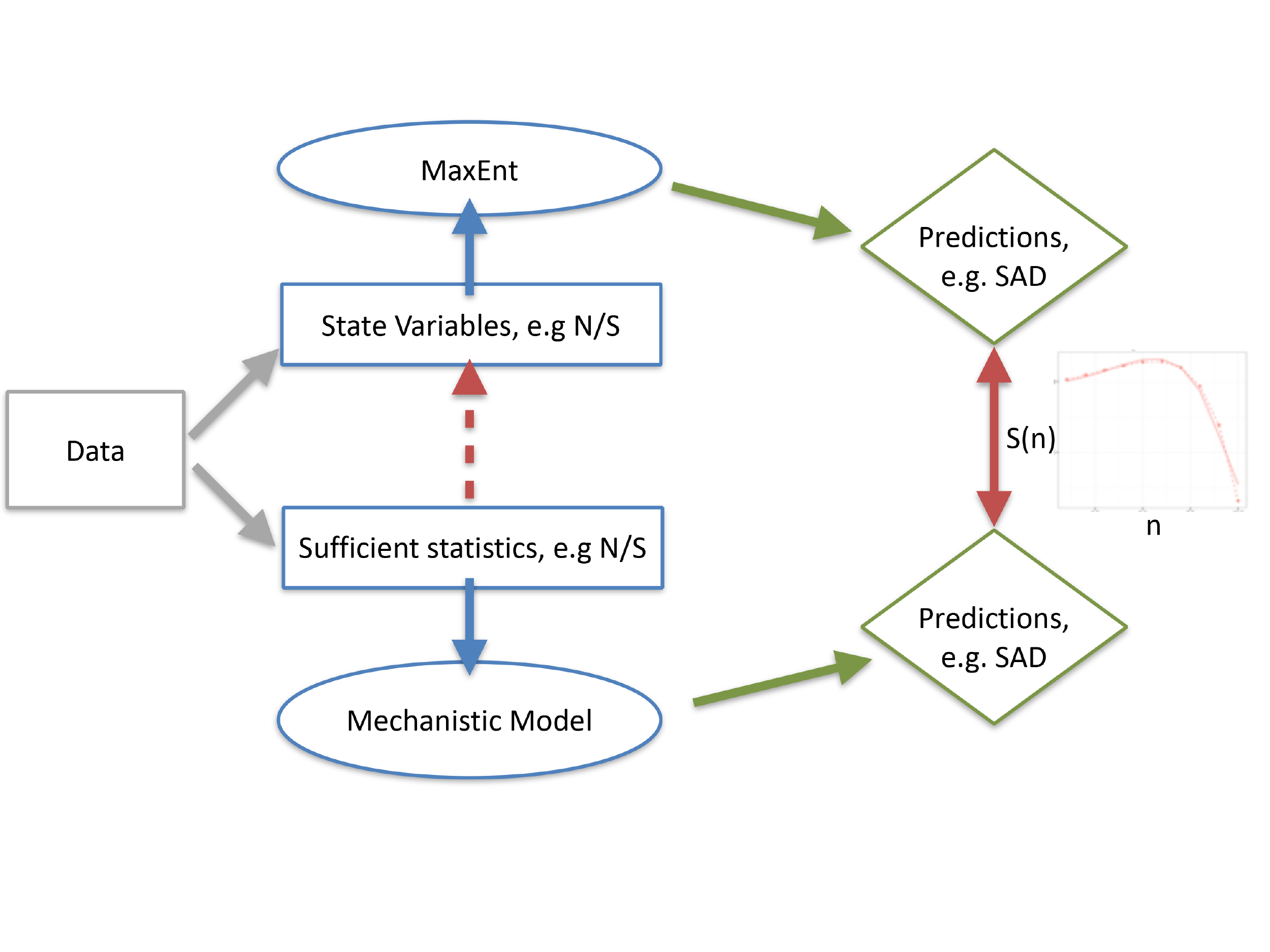}
\caption{\footnotesize  Both MaxEnt approaches and mechanistic models require input from empirical data (grey arrows). In the case of MaxEnt, this is a set of state variables. A common choice of state variable is the mean abundance per species, $N/S$, but in principle these state variables can be chosen however the modeler wishes.  In a given mechanistic model, the data overhead might be a finite set of sufficient statistics.  Both MaxEnt and mechanistic approaches take this specific input from data (blue arrows), and produce predictions for similar ecological patterns (green arrows).  A typical example would be the Species Abundance Distribution, which plots number of species ($S(n)$) as a function of abundance ($n$). The gap we identify is that these procedures have been carried out largely in parallel, with no direct connections between MaxEnt and mechanistic models. We propose that by specifying state variables using the sufficient statistics for a particular mechanistic model (dashed red line), we now put both approaches on the same footing, making the successes and failures of their predicted distributions directly comparable (solid red line). This can either provide a motivation for a particular set of state variables for MaxEnt, or can be used to provide a null comparison for a given mechanistic model.}
\label{fig:SSNT}
\end{figure}

\smallskip

\section{Reinterpreting MaxEnt as a Null Model}

The procedure of exactly how and which constraints should be chosen in existing MaxEnt approaches in ecology~\citep{shipley2006plant,pueyo2007,haegeman2008limitations,volkov2009inferring,harte2011maximum,harte2014maximum} is an open question. We propose that, \textbf{when constrained by the sufficient statistics of any given mechanistic model, MaxEnt can be used as a null hypothesis with which to test the value of the model}.  We show this proposal graphically in Figure~1. For every mechanistic model with predicted distributions belonging to an exponential family, we can identify its sufficient statistics as the constraints in what we call the `corresponding' MaxEnt theory.   If our mechanistic model can outperform its corresponding MaxEnt theory on a given data set, then specifying the details of the model and calculating its solution has been worthwhile. If not, whatever we have contributed to the construction of the model is only useful in so much as it fixes the constraints to measure using the data---beyond that, our efforts as modelers have been futile.  We propose the MaxEnt distribution as an appropriate null because the maximum entropy principle specifies as little as possible about the distribution beyond what is fixed by the sufficient statistics measured in a given data set.

To perform this comparison quantitatively, we propose an `entropically-corrected' likelihood, where for a given data set we take the likelihood of the mechanistic model, and subtract the likelihood of the corresponding MaxEnt distribution.  In the case of $S$ observations of a discrete variable $n$, and a mechanistic model with distribution $P_{\textrm{model}}(n)$ of the form given in Eq.~\eqref{eq:generalexpfam}, our proposed measure of performance takes the form:
\begin{align}
\log \mathcal{L}_{\textrm{corr}}(\textrm{model}| \{n_i   \})& =  \sum_{i=1}^{S} \log P_{\textrm{model}}(n_i|\alpha(\barF)) - \sum_{i=1}^{S} \log P_{\textrm{ME}}(n|\lambda(\barF))\nn\\
& = S\lt[\overline{\log(h)}+\barF\lt(\lambda(\barF)-\alpha(\barF)\rt)+\log \frac{A(\alpha(\barF))}{B(\lambda(\barF))}\rt].
\label{eq:entropiccorr}
\end{align}
where $P_{\textrm{ME}}(n|\lambda)$ is the MaxEnt distribution obtained by constraining the mean value of sufficient statistic $F(n)$. This is essentially applying analysis of Sections~1 and~2, and so $\alpha(\barF)$ and $\lambda(\barF)$ are given by Eqs.~\eqref{eq:generalmlest} and~\eqref{eq:fixlagrange}, respectively, while $A(\alpha)$ and $B(\lambda)$ are the corresponding normalizations of the mechanistic and MaxEnt distributions.

Drawing from the classic literature on exponential families~\citep{pitman1936sufficient,koopman1936distributions,darmois1945limites,jeffreys1960extension}, we note that the only possible difference between a mechanistic model distribution and its corresponding MaxEnt distribution arises in the form of the base measure, characterized as $h(n)$ above. We think of this as a model-implied base measure, and it leads to a reduction in entropy (relative to the uniform base measure) arising from our specification of the mechanism.  We note that the difference between $\alpha$ and $\lambda$ (with the same sufficient statistic, $F$), comes only from the fact that they have been estimated using different choices of $h(n)$. Our proposal is therefore  a kind of likelihood ratio test for whether  the model-implied base measure $h(n)$ provides a better explanation of our data than the uniform measure. Moreover, if we already have strong evidence for a particular base measure over the uniform measure~\citep{pueyo2007}, then we could also consider this as a new, more stringent null model for any new mechanistic prediction.  In other words, our approach can be extended to compare different sets of mechanisms, with the same sufficient statistics but different base measures $h(n)$.

\smallskip

\section{Applications to Empirical Data}

To provide a non-trivial mechanistic model, we turn to size-structured neutral theory (SSNT), and draw results below from~\citep{ODwyer2009,xiao2015comparing}, and the Supporting Information for this manuscript. This is an extension of the neutral ecological model introduced above, but with the addition of a new variable representing the size, mass, or energy flux of an individual. Speciation is defined in the same way as in the standard neutral theory, but now birth and death rates $b(m)$ and $d(m)$ can depend on the size or mass of an organism, $m$.  Also, there is a new process: ontogenetic growth. Each individual grows through time with a rate $g(m)$, which may also depend explicitly on its current mass. To fully specify this theory, we need to determine the functions $b(m)$, $d(m)$ and $g(m)$.  For this analysis, we parametrize these functions in the simplest way, by setting all three to be independent of mass, $m$, and we use the notation $b$, $d$ and $g$ for these three, constant rates. Even in this case, the combination of birth, death and growth still introduces variation in individual masses, as well as variation in the average size and total biomass across different species. 

The analysis of this section will provide an application of our approach using MaxEnt as a null model.  It also raises a new question.  For any given mechanistic model, there may be multiple possible distributions predicted, for example by marginalizing over some of the variables, which we could think of as unobserved.  Each of these different ways of formulating a predicted distribution then has its own corresponding MaxEnt. In the case of these size-structured neutral models, we highlight this by focusing on two cases, which we term coarse-grained and fine-grained.  In the coarse-grained prediction, we imagine we are only able to measure total biomass for each species, while in the fine-grained prediction we specify the biomasses of each individual.  Each of these has a different corresponding MaxEnt distribution, even though the constraints are the same, and below we explore the consequences of these differences.

\subsection{Size-Structured Neutral Theory: Coarse-grained Description}

First, we consider the joint distribution that a species chosen at random will have abundance $n$ and total biomass (summed across all $n$ individuals) $M$. Under the rules of SSNT, this distribution is~\citep{xiao2015comparing}:
\begin{equation}
P_{\textrm{SSNT}}(n,M) = \frac{1}{m_0\log(1/\nu)}\frac{(1-\nu)^n}{n!}\lt(\frac{M}{m_0}\rt)^{n-1}e^{-\frac{M}{m_0}}\label{eq:ssntjd}.
\end{equation}
where $n$ takes values in the positive integers and $M$ is a continuous variable $> 0$. (The latter definition is straightforward to generalize to account for a finite initial mass of new individuals). $\nu$ is the speciation rate in units of the generation time, while $m_0$ is a mass scale and is equal to the ratio of rates $g/d$.  Finally, we note that marginalizing over total biomass $M$ returns us to the simpler result for the log series species abundance distribution given in Eq.~\eqref{eq:neutrallogseries}. If one chooses not to measure species biomass, the predictions recapitulate the standard neutral theory.

\smallskip

The two sufficient statistics of the joint distribution $P_{\textrm{SSNT}}(n,M)$ given by Eq.~\eqref{eq:ssntjd} are mean biomass per species $\sum_i M_i /S = \overline{M}$ and mean abundance per species, $\sum_i n_i/S = \barn$. More explicitly, the maximum likelihood estimates of parameters $\nu$ and $m_0$ are given by:
\begin{align}
\frac{1-\nu}{\nu\log(1/\nu)}& = \barn\nn\\
m_0 = \frac{\barM}{\barn}\label{eq:ssntml}
\end{align}
We next carry out our strategy of constructing a MaxEnt distribution with uniform base measure to provide a baseline for the performance of $P_{\textrm{SSNT}}(n,M)$.  Constraining $\barM$ and $\barn$, we arrive at the following MaxEnt distribution for $M$ and $n$ in a size-structured community:
\begin{align}
P_{\textrm{SSME}}(n,M) = (e^{\lambda_1}-1)\lambda_2e^{-\lambda_1 n}e^{-\lambda_2M}
\end{align}
where the Lagrange multipliers impose the constraints on $\barM$ and $\barn$ and take the values:
\begin{align}
\lambda_1 & =\log\frac{\barn}{\barn-1} \nn\\
\lambda_2 & = 1/\barM
\end{align}
\blank{also fixed sign error here}
We now have an explicit, multivariate example of our proposed entropic correction, which takes the form:
\begin{align}
\log & \mathcal{L}_{\textrm{corr}}(\textrm{model}| \{n_i   \},\{M_{i}\}) \nn\\ & = \sum_{i=1}^S\lt(  \log\lt[\frac{1}{m_0\log(1/\nu)}\frac{(1-\nu)^{n_i}}{n_i!}\lt(\frac{M_i}{m_0}\rt)^{n_i-1}e^{-\frac{M_i}{m_0}}\rt]- \log\lt[(e^{\lambda_1}-1)\lambda_2e^{-\lambda_1 n_i}e^{-\lambda_2M_i}\rt]\rt)
\label{eq:ssntcorr}
\end{align}
where $n_i$ is the abundance of species $i$, and $M_{i}$ is the mass of species $i$, and the sum is over all $S$ observed species

In Figure~\ref{fig:SSNT} we use Eq.~\eqref{eq:ssntcorr} to evaluate the performance of the size-structured neutral theory (with  parameters set by Eq.~\eqref{eq:ssntml} and Lagrange multipliers also fixed using the data). For demonstration, we specifically examine two taxonomic groups with very different traits: trees and birds. We adopted forest plot data used in~\citep{xiao2015comparing}, all except for one to which we did not have access. These include 75 plots from four continents (Asia, Australia, North America, and South America), with 2189 species and morpho-species, and 380590 individuals in total~\citep{dryad_r9p70,Baribault2011,Bradford2014,Condit1998,Condit1998a,Condit2004,DeWalt1999,Gilbert2010,Hubbell2005,Hubbell1999,Kohyama2003,Kohyama2001,Lopez-Gonzalez2009,Lopez-Gonzalez2011,McDonald2002,Palmer2007,Peet1987,Pitman2005,Pyke2001,Ramesh2010,Reed1993,Thompson2002,Xi2008,Zimmerman1994}. All individuals have been identified to species or morpho-species, with measurement for diameter at breast height (DBH). We converted DBH to biomass using a metabolic scaling ansatz~\citep{west1999general,enquist2002global}. (For detailed description on the forest plot data and their manipulations, see~\citep{xiao2015strong,xiao2015comparing}. For a cleaned subset of these data, see the Dryad data package~\citep{dryad_5fn46,xiao2015strong}.)  For our second data set, we compiled all 2958 routes from the North American Breeding Bird Survey \citep{bbs} that were sampled during 2009.  These data are availible from US Geological Survey (\url{https://www.pwrc.usgs.gov/bbs/rawdata}).  Survey routes consist of 50 observation points, each sepparated by 0.5 mi. At each point all birds within 0.25 mi are identified and recorded by an expert observer.  Body size data were taken from \citep{dunning2007} and matched by taxonomy to records in the BBS data.  Both route data and body mass data are available at \url{https://github.com/ajrominger/MaxEntSuffStat}.

\begin{figure}
\setlength{\unitlength}{1cm}
\thicklines
\begin{picture}(10,8)
\put(0,0){\includegraphics[scale=0.65,clip=TRUE,trim= 0 0 00 0]{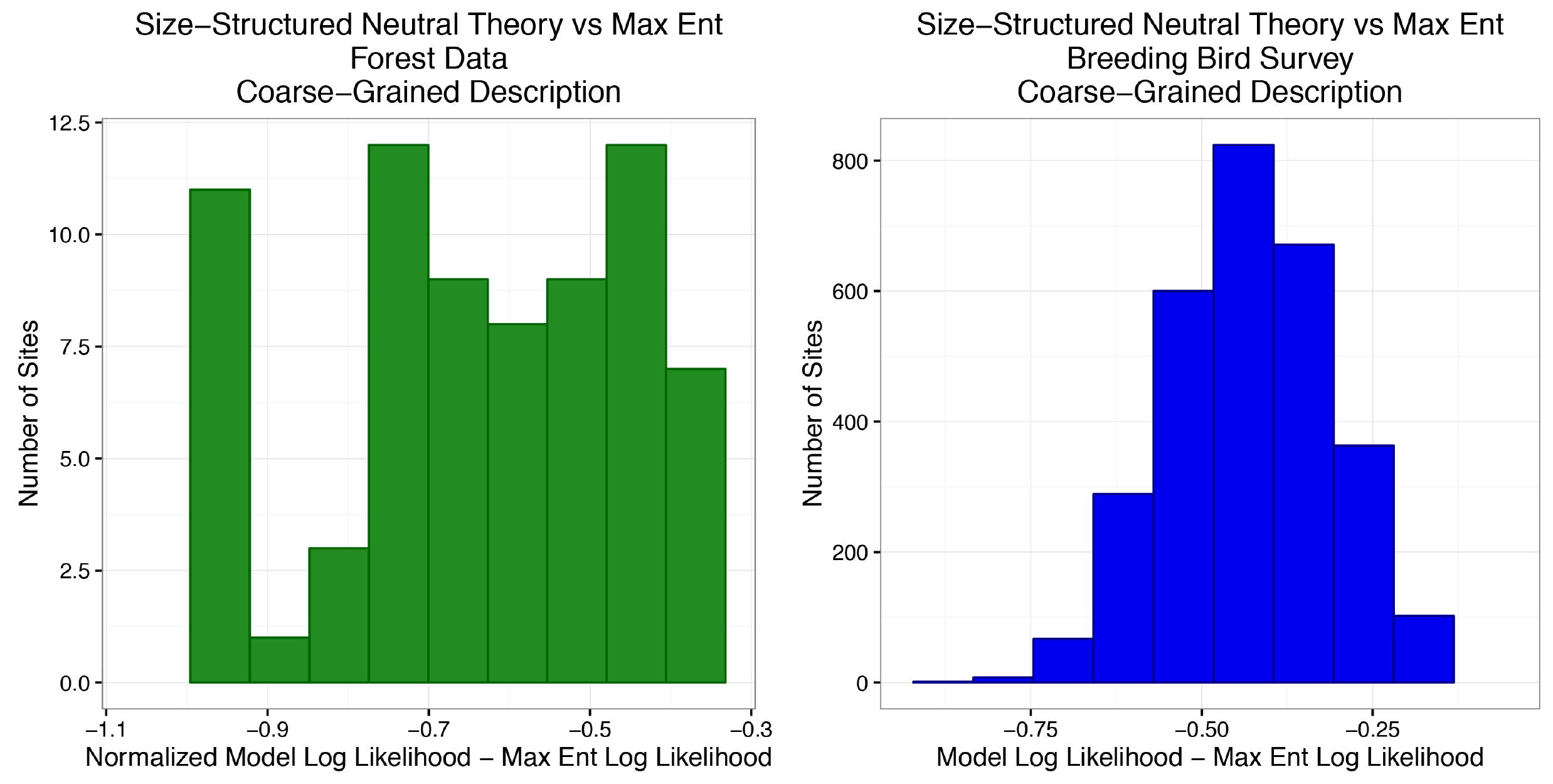}}
\end{picture}
\caption{\footnotesize  For 75 forest plots and 2958 locations from the Breeding Bird Survey, and using the `entropically-corrected' likelihood in Eq.~\eqref{eq:ssntcorr}, we test the performance of Size-Structured Neutral Theory against its corresponding Maximum Entropy theory.  In this figure we plot results for what we called the \textbf{coarse-grained description} in the main text, where only the total abundance and total biomass (or estimate of total biomass) is used to formulate predictions.  Negative values indicate that MaxEnt is a better description of the data than the neutral model, and positive values indicate that the neutral model provides a better description than MaxEnt. (The normalization here means that we divide this corrected log likelihood by the model log likelihood, so that we can put a wide range of likelihoods on the same plot---however, the normalization does not change the sign and hence interpretation.)  For all forest plots and all bird sites, we find that the size-structured neutral theory is a poor description compared to its MaxEnt counterpart.}
\label{fig:SSNT}
\end{figure}

Across these 75 forest plots and 2958 locations from the Breeding Bird Survey, we find a consistent result: in all locations, SSNT is outperformed by its MaxEnt baseline. In other words, if all you know about a forest plot or a bird community is its mean abundance per species $\barn$ and mean biomass per species $\barM$, we should almost always reject size-structured neutral dynamics as an explanation for its species abundance and biomass distributions.   In the case of the bird data, this maybe is unsurprising---a model with ontogenetic growth continuing througout an individual's lifetime will generate a broader range of intraspecific variation than we might expect in these species.  In the case of the forest data, it would have been less surprising for the neutral model to perform well, but we still find that SSNT performs worse than its corresponding MaxEnt distribution.  We note that while we interpret this as telling us that SSNT is a poor description of these data, it doesn't tell us that the corresponding MaxEnt distribution SSME is a good alternative.  In particular, since by design the constraints of SSME are identical to those of SSNT, the poor performance of SSNT may suggest that neither distribution (in absolute terms) is likely to be a good description of these data.

\subsection{Size-Structured Neutral Theory: Fine-grained Description}

We next consider a more fine-grained way to test the size-structured neutral theory.  In addition to measuring each species' abundance and its total biomass, we also measure the mass of each of its individuals. Replacing the joint distribution above for $n$ and $M$, we can make a neutral prediction for the precise distribution of masses within a species~\citep{xiao2015comparing}:
\begin{align}
P_{\textrm{SSNTI}}(n,m_1,\dots,m_n) & = \frac{1}{\log(1/\nu)}\frac{(1-\nu)^n}{n}\frac{1}{m_0^n}\prod_{j=1}^n e^{-m_j/m_0}\label{eq:ssnti}.
\end{align}
We have labeled this distribution `SSNTI', so that the I stands for individual-level.  The sufficient statistics for the parameters $\nu$ and $m_0$ are again given by mean abundance per species and mean total biomass per species:
\begin{align}
\frac{1-\nu}{\nu\log(1/\nu)}& = \barn\nn\\
m_0 = \frac{\overline{\sum_{j=1}^nm_j}}{\barn}\label{eq:ssntiml}.
\end{align}
Using these as constraints, we can in parallel construct the corresponding individual-level maximum entropy distribution to use as a baseline for the performance of $P_{\textrm{SSNTI}}(n,m_1,\dots,m_n)$:
\begin{align}
P_{\textrm{SSMEI}}(n,m_1,\dots,m_n) & =  (e^{\lambda_1}-1)\lambda_2^ne^{-\lambda_1 n}\prod_{j=1}^ne^{-\lambda_2 m_j}
\end{align}
where the Lagrange multipliers impose the constraints on $\overline{\sum_{j=1}^nm_j}$ and $\barn$ and take the values:
\begin{align}
\lambda_1 & =\log\frac{\barn}{\barn-1} \nn\\
\lambda_2 & = \frac{\barn}{\overline{\sum_{j=1}^nm_j}}
\end{align}
The corrected log likelihood for this case is then
\begin{align}
\log  \mathcal{L}_{\textrm{corr}}(\textrm{model}| \{n_i   \},\{m_{ij}\})& = \sum_i\lt(-\log \lt[(e^{\lambda_1}-1)\log(1/\nu)\rt] +n_i\log\lt[e^{\lambda_1}(1-\nu)\rt]-\log[{n_i}] \rt)\label{eq:ssnticorr}
\end{align}
where $m_{ij}$ is the mass of the $j$-th individual from species $i$.  In fact, in this expression all of the mass dependence cancels from the two terms, leaving us with the comparison of a log series and geometric series. The mathematical independence of this quantity on individual masses allows us to calculate it even for the breeding bird data, for which no individual mass estimates are available.

\begin{figure}
\setlength{\unitlength}{1cm}
\thicklines
\begin{picture}(10,8)
\put(0,0){\includegraphics[scale=0.65,clip=TRUE,trim= 0 0 00 0]{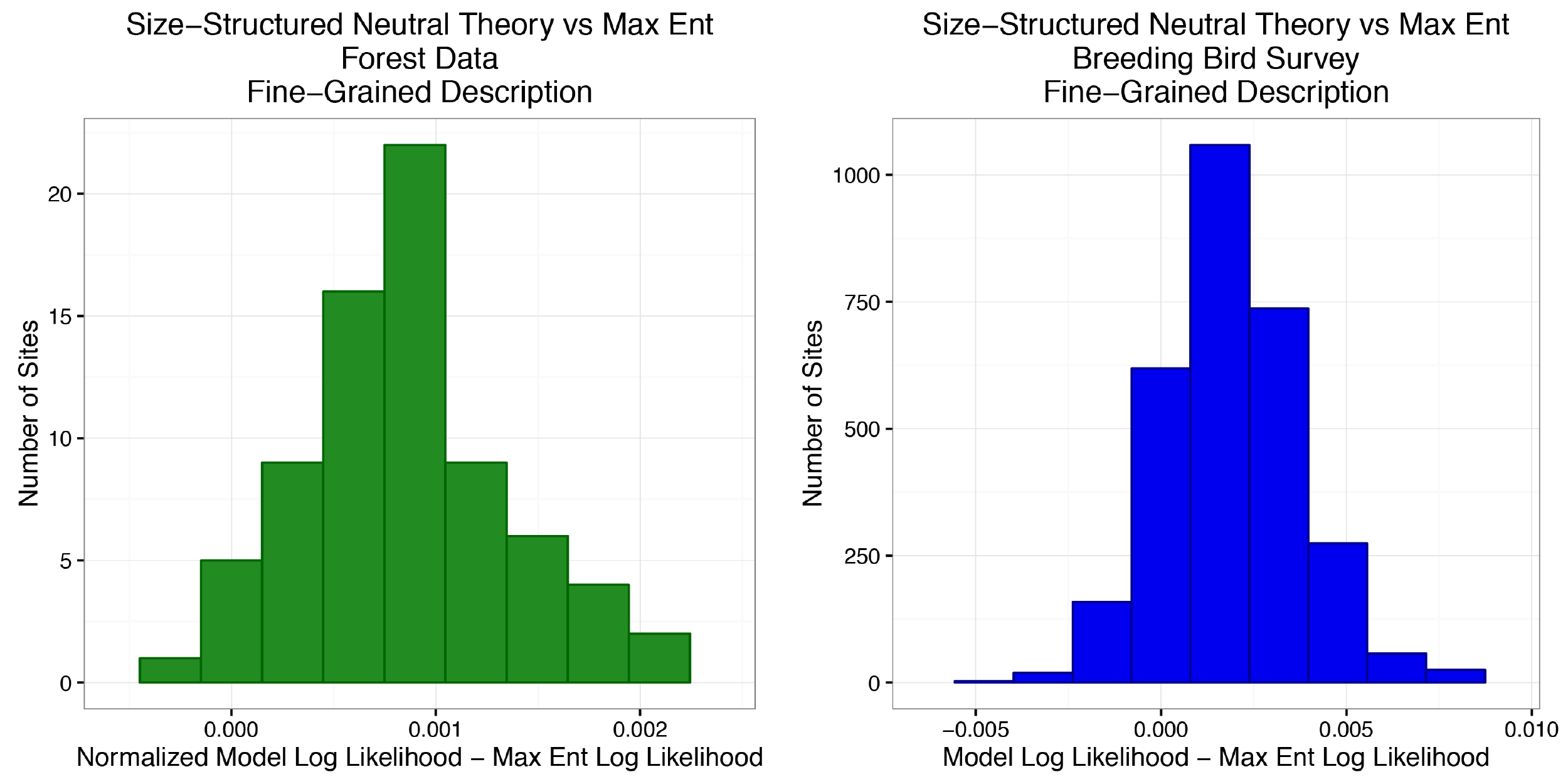}}
\end{picture}
\caption{\footnotesize For 75 forest plots and 2958 locations from the breeding bird survey, we test the performance of Size-Structured Neutral Theory against its corresponding Maximum Entropy theory, with constraints determined from the sufficient statistics of the neutral theory.  In this figure we plot results for what we called the \textbf{fine-grained description} in the main text, where  the total abundance and each stem or individual biomass is used to formulate predictions, though in practice Eq~\eqref{eq:ssnticorr} only depends on species abundances.  Negative values indicate that MaxEnt is a better description of the data than the neutral model, and positive values indicate that the neutral model provides a better description than MaxEnt. (The normalization here means that we divide this corrected log likelihood by the model log likelihood, so that we can put a wide range of likelihoods on the same plot---however, the normalization does not change the sign and hence intepretation.) For the majority of forest plots and bird sites, we find that neutral theory is a better description than its MaxEnt counterpart. We are able to evaluate this comparison for Breeding Bird Survey data despite lacking individual variation in body size/mass, because the comparison of the two log-likelihoods reduces to a function that depends only on species abundances, as shown in Eq.~\eqref{eq:ssnticorr}.  The log series component of the neutral prediction is therefore a better description of species abundances in these data than the geometric series component of the corresponding MaxEnt prediction.}
\label{fig:SSNTI}
\end{figure}

In Figure~\ref{fig:SSNTI} we evaluate the performance of the individual-based size-structured neutral theory by computing its log likelihood (with  parameters set by Eq.~\eqref{eq:ssntml}), with a maximum entropy baseline given by $P_{\textrm{SSMEI}}(n,m_1,\dots,m_n)$, with Lagrange multipliers fixed using the data.   Across the same forest and breeding bird plots as shown in Figure~\ref{fig:SSNT}, we see that SSNTI is almost universally a better explanation of the data relative to the corresponding maximum entropy distribution, as it is for the breeding bird data.  What changed? The individual-based SSNTI neutral model has a larger number of independent variables than its aggregated counterpart SSNT, but when conditioned on a fixed total biomass for a species, $\sum_{j=1}^nm_j=M$, $P_{\textrm{SSNTI}}(n,m_1,\dots,m_n)$ becomes equal to $P_{\textrm{SSNT}}(n,M)$: if you blur your eyes and only pick up on total biomass, the two neutral predictions are identical, as they should be. The same is not true of the two MaxEnt distributions, labeled SSME and SSMEI.  What changed is that we implicitly told $P_{\textrm{SSMEI}}$ that total biomass $M$ is comprised of a set of individuals of masses $\{m_j\}$.  The result is that the SSMEI model is identical to the SSNTI distribution in terms of the biomass factor, but differs in its prediction of the species abundance distribution.  So all we are seeing in Figure~\ref{fig:SSNTI} is that the classic log series SAD is generally a better description for these data than the geometric distribution.   It is not clear to us whether the similarity between the biomass terms in MaxEnt and the mechanistic model here is a general consequence of the increase in the degrees of freedom (here going from SSNT to the more fine-grained SSNTI) or if it is special to this case.  Further more systematic investigation of the relationship between MaxEnt and mechanistic distributions as a function of aggregating degrees of freedom may be the most appropriate strategy to clarify this issue.

\section{Discussion}

In this manuscript we related biological mechanism to the constraints used in the Maximum Entropy (MaxEnt) approach to predicting macroecological patterns. We achieved this by proposing that the sufficient statistics of a mechanistic model should be used as MaxEnt constraints, but the procedure we  introduced is incapable on its own of identifying a unique set of constraints for MaxEnt. Instead, we have (potentially) a different MaxEnt prediction corresponding to each different set of mechanisms, and we proposed that the natural way to use this prediction is as a null hypothesis. This null hypothesis has the properties of specifying unambiguously what quantities should be constrained, and it also does not require or invoke any alternative mechanism for comparison~\citep{gotelli2006null,connor1979assembly,harvey1983null,gotelli2006neutral}. In a sense, our null hypothesis is obtained by removing from the mechanistic distribution all mechanisms that create a bias over the support of the distribution, but retaining the aspects of mechanisms that defined the support in the first place. We propose that if a mechanistic model performs worse in a given data set than its corresponding MaxEnt distribution, then this provides evidence against the mechanisms and assumptions of the model. 

We demonstrated this by testing size-structured neutral models against their corresponding MaxEnt baselines, using empirical data drawn from multiple forest plots and the Breeding Bird Survey.  This test raised another question: how fine-grained is our description of the data, and consequently how many degrees of freedom are there in our model's predicted probability distribution?  For example, in this case of forest data, we may be able to estimate just total species biomass, or we may measure each individual stem. In this analysis, we found that whether mechanistic distributions were favored over MaxEnt or vice versa depended not only on the mechanism, but also on the number of degrees of freedom used to describe the data.  MaxEnt was generally favored when describing data in terms of total species biomasses, while the size-structured neutral theory was favored when describing data in terms of individual masses. However, our analysis does not clarify whether in general there will be a systematic relationship between mechanistic model success/failure in these terms as we aggregate more  or fewer degrees of freedom.

Where does this leave the Maximum Entropy Theory of Ecology~\citep{harte2011maximum} (METE), which prescribes a particular set of constraints, and makes predictions of the same types of distributions as the above?  In previous work the results of  METE have been compared with e.g. neutral models~\citep{xiao2015comparing}.  But the MaxEnt distributions derived in this paper were specifically chosen to match the sufficient statistics of a given mechanistic model, and do not precisely match the distributions predicted by METE. In fact, we do not know of any flavor of mechanistic theory whose independent variables and sufficient statistics precisely match the standard METE degrees of freedom and state variables, but we note that the sufficient statistics of the size-structured neutral models, namely involving average species abundance and biomass, are extremely close to the METE state variables.  Clarifying exactly what ranges of ecological mechanisms lead to these sufficient statistics might help us to understand why the METE state variables seem to perform well in the cases that they do, and might also give us insight into where METE might be expected to break down. Moreover, if we were able to show that certain sets of sufficient statistics are more likely than any others when looking across a range of ecological and evolutionary mechanisms, this would open the door to established preferred sets of state variables in a principled way.

Several important caveats in our approach are worth emphasizing. First, our example mechanistic models have (i) a finite set of sufficient statistics, (ii) the dimension of this set does not increase with sample size, and (iii) the support of the predicted distribution does not vary with parameter values. These features meant that both maximum entropy distributions and mechanistic model distributions belonged to an exponential family, as defined in Section~1. Not all interesting mechanistic models in ecology will share these features, as many commonly-predicted probability distributions do not belong to exponential families. Second, we have assumed that model parameter values are either known, and fixed independently of a dataset, or are free parameters to be estimated using the current data, and have not tackled intermediate cases where we have partial knowledge of these parameters. Third,  our approach does not tell us if either a mechanistic model or its MaxEnt counterpart are good descriptions of the data in absolute terms. For example, if there are too many constraints, apparently good fits of a given mechanistic model or its corresponding MaxEnt may still be uninformative~\citep{haegeman2008limitations,shipley2009limitations}. I.e. our approach does not evaluate whether either of these distributions is overfitting a given data set. Finally, we focused on steady-state predictions. On the other hand, the prediction of fluctuation sizes on various timescales is precisely where simplified mechanistic models seem to break down~\citep{Chisholm2014b,chisholm2014ages,odwyer2015phylo,fung2016reproducing}.  At this point, we do not have a corresponding maximum entropy baseline for these models.

%% Enter the text of your article beginning here and ending before
%% \begin{acknowledgements}
%% Section head commands for your reference:
%% \section{}
%% \subsection{}
%% \subsubsection{}

\section*{Acknowledgments}

We thank three reviewers for an excellent and constructive set of reviews, which helped to shape and convey the main messages of this manuscript. We also acknowledge extensive and helpful feedback from Cosma Shalizi and Ethan White on earlier drafts of the manuscript. JOD acknowledges the Simons Foundation Grant \#376199, McDonnell Foundation Grant \#220020439, and Templeton World Charity Foundation Grant \#TWCF0079/AB47. AJR acknowledges funding from NSF grant DEB \#1241253. R. K. Peet provided data for the North Carolina forest plots. T. Kohyama provided the Serimbu dataset through the PlotNet Forest Database. The eno-2  plot (by N. Pitman) and DeWalt Bolivia (by S. DeWalt) datasets were obtained from SALVIAS. The BCI forest dynamics research project was made possible by NSF grants to S. P. Hubbell: DEB \#0640386, DEB \#0425651, DEB \#0346488, DEB \#0129874, DEB \#00753102, DEB \#9909347, DEB \#9615226, DEB \#9405933, DEB \#9221033, DEB \#-9100058, DEB \#8906869, DEB \#8605042, DEB \#8206992, DEB \#7922197, support from CTFS, the Smithsonian Tropical Research Institute, the John D. and Catherine T. MacArthur Foundation, the Mellon Foundation, the Small World Institute Fund, and numerous private individuals, and through the hard work of over 100 people from 10 countries over the past two decades. The UCSC Forest Ecology Research Plot was made possible by NSF grants to G. S. Gilbert (DEB \#0515520 and DEB \#084259), by the Pepper-Giberson Chair Fund, the University of California, and the hard work of dozens of UCSC students. These two projects are part CTFS, a global network of large-scale demographic tree plots. The Luquillo Experimental Forest Long-Term Ecological Research Program was supported by grants BSR \#8811902, DEB \#9411973, DEB \#0080538, DEB \#0218039, DEB \#0620910 and DEB \#0963447 from NSF to the Institute for Tropical Ecosystem Studies, University of Puerto Rico, and to the International Institute of Tropical Forestry USDA Forest Service, as part of the Luquillo Long-Term Ecological Research Program. Funds were contributed for the 2000 census by the Andrew Mellon foundation and by CTFS. The U.S. Forest Service and the University of Puerto Rico gave additional support. We also thank the many technicians, volunteers and interns who have contributed to data collection in the field.

\pagebreak

\begin{flushleft}
{\Large
\textbf{Supplementary Material}
}
% Insert Author names, affiliations and corresponding author email.
\\
James P. O'Dwyer $^{1}$, 
Andrew Rominger $^{2}$, 
Xiao Xiao $^3$
\\
\bf{1} Department of Plant Biology, University of Illinois, Urbana IL USA \\
\bf{2} Department of Environmental Science, Policy and Management, University of California, Berkeley, USA \\
\bf{3} School of Biology and Ecology, and Senator George J. Mitchell Center for Sustainability Solutions, University of Maine, Orono ME USA\\
\end{flushleft}
\appendix
\section{Derivation of Size-Structured Neutral Theory Results}
In~\citep{ODwyer2009}, we derived an exact solution  for a population undergoing birth at a rate $b$, mortality at a rate $d(m)$,  growth rate $g(m)$, and immigration rate $\nu$.  We expressed the solution in terms of a generating functional, $Z[H(m)]$, which is formulated as the limiting case of 
\begin{equation}
\mathcal{Z}[\{h_i\}] = \sum_{\{h_i\}} P(\{n_i\})e^{\sum_ih_in_i}\label{eq:discretedef}
\end{equation}
as a set of discrete size classes labeled by $i$ becomes a continuum. In this discrete case, $P(\{n_i\})$ is the probability that the population has $n_i$ individuals in size class $i$. The community-level interpretation of this is as the probability of a given species, chosen at random from a neutral community, having a set of individuals with different sizes $n_i$. For simplicity, we consider the case where the mass of the smallest individuals is infinitesmally small, though this can be generalized. This leads to the following solution for the size-structured neutral theory generating functional:
\begin{equation}
\mathcal{Z}[H(m)] =1- \frac{\log \lt[1-\log(1/\nu)\int_{0}^{\infty} dm f(m)( e^{H(m)}-1)\rt]}{\log\lt[1+\log(1/\nu)\int_{0}^{\infty} dm f(m)\rt]}.\label{eq:soln3}
\end{equation}
Note that the form of this result differs slightly from~\citep{ODwyer2009}. In keeping with our other versions of neutral models in this paper, we have defined speciation rate here to be a dimensionless per capita speciation rate, in units of the birth rate, $b$. We are also conditioning on abundances $n>0$ (in~\citep{ODwyer2009} we considered a formulation which kept track of a class of extinct species with $n=0$, and we have removed this here).

 Growth and mortality rates are then encoded in the function $f(m)$, which satisfies:
\begin{align}
\frac{d}{dm}\lt(g(m)f(m)\rt)+d(m)f(m)=0\label{eq:demeq3}\\
f(0)g(0) = -\frac{b}{\log\nu} +b\int_{0}^{\infty} dm\ f(m)\label{eq:dembound3}.
\end{align}
From this generating functional, we can obtain the generating functions for various joint probability distributions using particular functional forms for the auxiliary function, $H(m)$. In~\citep{ODwyer2009}, we solved for the species abundance distribution by setting $H(m) = h_0$, and for the species biomass distribution by setting $H(m)=h_1m$. These relationships follow from the limit of the definition~Eq.~\eqref{eq:discretedef}.

\subsection{Coarse-grained case}

To obtain the generating function for the joint distribution of total abundance and total biomass, we correspondingly need to set $H(m) = h_0 + h_1m$ in Eq.~\eqref{eq:soln3}. This gives:
\begin{equation}
z(h_0,h_1)=   1- \frac{\log \lt[1-\log(1/\nu)\int_{0}^{\infty} dm f(m)( e^{h_0+h_1m}-1)\rt]}{\log\lt[1+\log(1/\nu)\int_{0}^{\infty} dm f(m)\rt]}.\label{eq:soln4}
\end{equation}
This generating function can then be transformed back into the following probability distribution for $P(n,M)$:
\begin{equation}
P(n,M) = \frac{1}{n}\lt(\frac{\log[1/\nu]}{1+\log(1/\nu)\int_{0}^{\infty} dm f(m)}\rt)^n \underbrace{f\star\dots\star f(M)}_{n\ \textrm{times}}
\end{equation}
where the biomass dependence is in the form of a product of multiple convolutions.  This can be checked by direct substitution:
\begin{align}
\int_{0}^\infty dM\sum_{n=1}^\infty P(n,M)e^{h_0n+h_1M} & = \sum_{n=1}^\infty e^{h_0n}\frac{1}{n}\lt(\frac{\log[1/\nu]}{1+\log(1/\nu)\int_{0}^{\infty} dm f(m)}\rt)^n \int_{0}^\infty dMe^{h_1M}\underbrace{f\star\dots\star f(M)}_{n\ \textrm{times}}\nn\\
& = \sum_{n=1}^\infty e^{h_0n}\frac{1}{n}\lt(\frac{\log[1/\nu]}{1+\log(1/\nu)\int_{0}^{\infty} dm f(m)}\rt)^n\lt[\int_{0}^\infty f(M)e^{h_1M} dM\rt]^n\nn\\
& = -\frac{\log \lt[1-\frac{\log[1/\nu]\int_{0}^{\infty}dM f(M) e^{h_0+h_1M}}{1+\log[1/\nu]\int_{0}^{\infty}dM f(M) }\rt]}{\log\lt[1+\log(1/\nu)\int_{0}^{\infty} dm f(m)\rt]}\nn\\
& = 1- \frac{\log \lt[1-\log(1/\nu)\int_{0}^{\infty} dm f(m)( e^{h_0+h_1m}-1)\rt]}{\log\lt[1+\log(1/\nu)\int_{0}^{\infty} dm f(m)\rt]}
\end{align}
This result is general, and can be applied in cases where $g(m)$ and $d(m)$ depend on individual body size. In this paper, we focused instead on the `completely neutral' limit, where in fact individuals have identical rates $b$, $d$ and $g$ independent of their size/mass.  In this case, we had shown earlier~\citep{ODwyer2009} that solving Eq.~\eqref{eq:dembound3} (again adapting to the per capita definition of $\nu = 1-b/d$ that we use throughout this current paper) results in:
\begin{align}
f(m) =\frac{1-\nu}{\nu\log(1/\nu)} \frac{d}{g}e^{-\frac{d}{g}m}
\end{align}
Note that when we integrate over all sizes, we find:
\begin{align}
\int_0^{\infty} f(m) = \frac{1-\nu}{\nu\log(1/\nu)},
\end{align}
which is the standard non-zero sum neutral theory result for the total number of individuals divided by the expected number of species.  Hence we have (when integrated over all size classes) the correct expression for the total number of individuals per species.

The exponential function belongs to the larger class of Gamma distributions, which in turn is a particular case of a Tweedie distribution. Tweedie distributions have the nice property that we can convolve them with themselves as many times as we like, and the result takes the same functional form but with rescaled parameters.  This makes computing the convolution product straightforward, and for this case we have:
\begin{align}
\underbrace{f\star\dots\star f(M)}_{n\ \textrm{times}} =\frac{1}{(n-1)!}\lt( \frac{1-\nu}{\nu\log(1/\nu)}\rt)^n \frac{d}{g}\lt(dM/g\rt)^{n-1}e^{-\frac{d}{g}M}.
\end{align}
Putting this together with the general result above, we have for the `coarse-grained' size-structured neutral theory:
\begin{align}
P(n,M) & =  \frac{1}{n!}\lt(\frac{\log[1/\nu]}{1+ \frac{1-\nu}{\nu}}\rt)^n \lt( \frac{1-\nu}{\nu\log(1/\nu)}\rt)^n \frac{d}{g}\lt(dM/g\rt)^{n-1}e^{-\frac{d}{g}M}\nn\\
& = \frac{1}{n!} \lt( 1-\nu\rt)^n \frac{d}{g}\lt(dM/g\rt)^{n-1}e^{-\frac{d}{g}M}\end{align}
This is what we reported in the main text, where we defined a size/mass scale $m_0=g/d$ for notational convenience (note that this is distinct from the notation used in~\citep{ODwyer2009}, where $m_0$ was used to denote the minimum mass of an individual).

\subsection{Fine-grained case}
To obtain the generating function for the joint distribution of total abundance and all individual biomasses for the completely neutral size-structured theory, we first consider the distribution of individual biomasses conditioned on total abundance being $=n$.  The generating function of this distribution can be identified by treating the auxiliary function as a constant term plus an additionla function, $H(m) = h_0 + h(m)$, expanding $\mathcal{Z}[H(m)]$ in powers of $e^{h_0}$, and extracting the term proportional to $e^h_0n$, to obtain:
\begin{align}
Z[h(m)] =\lt(\frac{\int_{0}^{\infty} dm f(m)e^{h(m)}}{\int_{0}^{\infty} dm f(m)}\rt)^n\label{eq:soln6}
\end{align}
We also note that when conditioned on $\int_0^\infty n(m) = n$, the only allowable size-spectra must take the form
\begin{align}
n(m) = \sum_{i=1}^n \delta(m-m_i)
\end{align}
where $m_i$ is the mass of individual $i$, and we have used the Dirac delta function.  I.e. the spectrum of a species with exactly $n$ individuals must at any one point in time consist of a set of infinitely-sharp spikes located at the masses of its constituent individuals. Hence we can write:
\begin{align}
Z[h(m)] = \int [dn] \mathcal{P}[n(m)|n] e^{\int_0^\infty h(m)n(m)} = \int \prod_{i=1}^n dm_i P_{\textrm{SSNTI}}(\{m_i\}|n)e^{\sum_{i=1}^n h(m_i)}.
\end{align}
where $\mathcal{P}[n(m)|n] $ is the a functional giving the probability of a species consisting of a size/mass spectrum $n(m)$ when conditioned on total abundance $n$, while $P_{\textrm{SSNTI}}(\{m_i\}|n)$ is an equivalent description in terms of the probability that the same species consists of $n$ individuals with the specific set of $n$ biomasses $\{m_i\}$.  From the form of Eq.~\eqref{eq:soln6} we then have 
\begin{equation}
P_{\textrm{SSNTI}}(\{m_i\}|n) = \prod_{i=1}^n \frac{f(m_i)}{\int_{0}^{\infty} dm f(m)}
\end{equation}
In the completely neutral case, $\frac{f(m_i)}{\int_{0}^{\infty} dm f(m)} = \frac{d}{g}e^{-\frac{d}{g}m_i}$, and also $P_{\textrm{NT}}(n) = \frac{1}{n\log(1/\nu)}{(1-\nu)^n}$, and putting these results together gives us that:
\begin{align}
P_{\textrm{SSNTI}}(n,\{m_i\})& =P_{\textrm{SSNTI}}(\{m_i\}|n)P_{\textrm{NT}}(n) \nn\\
& =\frac{1}{\log(1/\nu)}\frac{(1-\nu)^n}{n}\frac{1}{m_0^n}\prod_{j=1}^n e^{-m_j/m_0}
\end{align}
where $m_0=g/d$ as in the main text.

%% == end of paper:

%% Optional Materials and Methods Section
%% The Materials and Methods section header will be added automatically.

%% Enter any subheads and the Materials and Methods text below.

%% Optional Appendix or Appendices
%% \appendix Appendix text...
%% or, for appendix with title, use square brackets:

%%
%\bibliographystyle{ecology_letters2}
%\bibliography{maxentrefs}{}
%additional bib files
%\bibliography{andyrefs}
%\bibliography{xiaodata.bib}
%}

%\end{article}

%%%%%%%%%%%%%%%%%%%%%%%%%%%%%%%%%%%%%%%%%%%%%%%%%%%%%%%%%%%%%%%%

%% Adding Figure and Table References
%% Be sure to add figures and tables after \end{article}
%% and before \end{document}

%% For figures, put the caption below the illustration.
%%
%% \begin{figure}
%% \caption{Almost Sharp Front}\label{afoto}
%% \end{figure}

%% For Tables, put caption above table
%%
%% Table caption should start with a capital letter, continue with lower case
%% and not have a period at the end
%% Using @{\vrule height ?? depth ?? width0pt} in the tabular preamble will
%% keep that much space between every line in the table.

%% \begin{table}
%% \caption{Repeat length of longer allele by age of onset class}
%% \begin{tabular}{@{\vrule height 10.5pt depth4pt  width0pt}lrcccc}
%% table text
%% \end{tabular}
%% \end{table}

%% For two column figures and tables, use the following:

%% \begin{figure*}
%% \caption{Almost Sharp Front}\label{afoto}
%% \end{figure*}

%% \begin{table*}
%% \caption{Repeat length of longer allele by age of onset class}
%% \begin{tabular}{ccc}
%% table text
%% \end{tabular}
%% \end{table*}

\end{document}